\documentclass[12pt]{iopart}
\usepackage{graphicx}
\usepackage{amssymb}

\begin{document}

%
\title{Spontaneous reorientations in a model of opinion dynamics with
anticonformists}
\author{Grzegorz Kondrat and Katarzyna Sznajd-Weron}

\address{Institute of Theoretical Physics, University of Wroc{\l}aw, Pl. Maxa Borna 9,
50-204 Wroclaw, Poland}
\ead{kweron@ift.uni.wroc.pl}
\begin{abstract}
In this paper we investigate a model (based on the idea of the outflow dynamics), in which only conformity and anticonformity can lead to the opinion change. We show that for low level of aniconformity the consensus is still reachable but spontaneous reorientations between two types of consensus ('all say yes' or 'all say now') appear. 
\end{abstract}
\vspace{2pc}

\maketitle

\section{Introduction}
In the past decade many models of opinion dynamics has been studied by physicists (for the recent review see \cite{Castellano_2009}). Among them several simple discrete models based on the famous Ising model, such as Voter model \cite{Liggett_1999}, majority models \cite{Galam_2002,Krapivsky_2003} or Sznajd model \cite{Sznajd_2000}, have been proposed to describe consensus formation. The force which leads to consensus is conformity -- one of the most observed response to the social influence. In all three models mentioned above a kind of conformity has been introduced. In the Voter model a single person is able to convince others, within the majority rule individuals follow majority opinion and in the Sznajd model unanimity is needed to convince others. Although the conformity is the major paradigm of the social influence, it is known that other types of social response are also possible. 

People feel uncomfortable when they appear too different from others, but they also feel uncomfortable when they appear like everyone else \cite{Myers_1996}. There is an experimental evidence for asserting uniqueness - sometimes people to assert their uniqueness can change their own opinion, when they realize that this opinion is shared by others \cite{Myers_1996}. Therefore asserting uniqueness can lead to so called anticonformity. In 1963 Willis (reviewed recently in \cite{Nail_2000}) has proposed a two-dimensional model of possible responses to social influence, in which both conformers and anticonformers are similar in the sense that both acknowledge the group norm (the conformers agree with the norm, the anticonformers disagree). 


Obviously the anticonformity is quite rare in comparison to the conformity. The natural question is whether the existence of the very small probability of anticonformity can influence the opinion dynamics. Will the consensus be still possible in the society with anticonformists? In this paper we decided to introduce the probability of anticonformal behavior to one of the consensus models. Recently a generalized one-dimensional model based on the original Sznajd model has been proposed to  incorporate some diversity or randomness in human activity \cite{Kondrat_2009}. In this paper we investigate a special case of this extended model, in which both conformity and anticonformity are possible. We check how the small probability of anticoformal behavior in the presence of the strong conformity can influence the opinion dynamics. It has been known for long that conformity/anticonformity  is to some extent a product of cultural conditions \cite{Bond_1996}. There are some experimental motivations for such statement. For example, Frager in 1970  conducted experiments among Japanese students and found a lower level of conformity compared with the U.S. results and some evidence for anticonformity \cite{Frager_1970}. From this point of view a ratio between the probability of conformity and anticonformity could be related to the cultural or political conditions. 

\section{The model}
We consider a chain of $L$ Ising spins $S_i=\pm 1, \; i=1,\ldots,L$ with periodic boundary conditions. At each step two consecutive spins are chosen at random, and they influence their outer neighbors. In the most popular version of the Sznajd model, inspired by the observation that an individual who breaks the unanimity principle reduces the social pressure of the group dramatically \cite{Myers_1996}, only the unanimous majority influences the neighborhood. In the paper \cite{Kondrat_2009} all possible configurations of 4 consecutive spins has been considered. Two randomly selected middle spins decide the outcome of the update step (following \cite{Kondrat_2009} we write them in brackets). 
The action of a selected pair has been considered independently on each direction. Thus all different possible elementary cases make up a following list:
($[AA]A$, $[AA]B$, $[AB]A$ and $[AB]B$), where the symbols $A$ and $B$ stand for different opinions, i.e $A=-B=\pm 1$. 
To determine the dynamics the vector of probabilities ${\bf p}=(p_1,p_2,p_3,p_4)$ 
of change the third spin (one that is outside brackets) has been introduced \cite{Kondrat_2009}:
\begin{eqnarray}
p_1:[AA]A \rightarrow [AA]B,\\ 
p_2:[AA]B \rightarrow [AA]A,\\ 
p_3:[AB]A \rightarrow [AB]B,\\ 
p_4:[AB]B \rightarrow [AB]A.
\end{eqnarray}
The first parameter, $p_1$, describes the chance of spontaneous appearing an anticonformist
opinion and the complementary probability $p_1'=1-p_1$ describes the situation, where in the same conditions 
the opinion is not changed.
Second parameter, $p_2$, is a chance of convincing an individual to the other opinion, 
shared by his two consecutive neighbours - i.e. conformity. Again $p_2'=1-p_2$ is a probability of one's opinion
remaining unaltered with the presence of conformity among his two consecutive neighbors
In this paper we investigate the special case, in which only conformity and anticonformity can lead to the opinion change, thus $p_3=p_4=0$. The case in which $p_2=1$ and $p_1=p_3=p_4=0$ corresponds to the Sznajd model. In this paper we have decided to investigate the case in which $p_2=1$ and $p_1 \in (0,1) $ is the only parameter of the model. To investigate the model, we provide Monte Carlo simulations with the random sequential updating mode and thus the time $t$ is measured in the Monte Carlo Steps (MCS) which consists of $L$ elementary updatings. 

\section{Results}
The quantity, which is usually measured in such models, is the public opinion $m$ as a function of time $t$. In this kind of models the public opinion is equivalent to the magnetization:
\begin{equation}
m=\frac{1}{L} \sum_{i=1}^N S_i.
\end{equation}
In the case of $p_1=0$, which corresponds to the deterministic rule of the Sznajd model, the system reaches the ferromagnetic steady state (consensus from the social point of view). Once $p_1>0$ the system never reaches any absorbing state and the opinion dynamics depends on anticonformity probability $p_1$. The time evolution of public opinion $m(t)$ is presented in Figs. 1-3. It can be seen that consensus ($m = \pm 1$) is reached only for small values of $p_1$ (Fig.1), while for larger values of anticonformity consensus is not reached and public opinion fluctuates around its mean value $m=0$ (Figs.2-3). One can also notice that the amplitude of the fluctuations decrease with $p_1$, on the other hand the frequency of fluctuations increase with $p_1$. This tendency is valid for all values of $p_1$ and thus the time of consensus state ('all up' or 'all down') decreases with $p_1$. For very small values of $p_1$ the system spends most of the time in one of the extreme consensus state and in the limiting case $p_1=0$ the consensus becomes the absorbing steady state.

\begin{figure}
\begin{center}
\vspace*{10mm}
\includegraphics[scale=1]{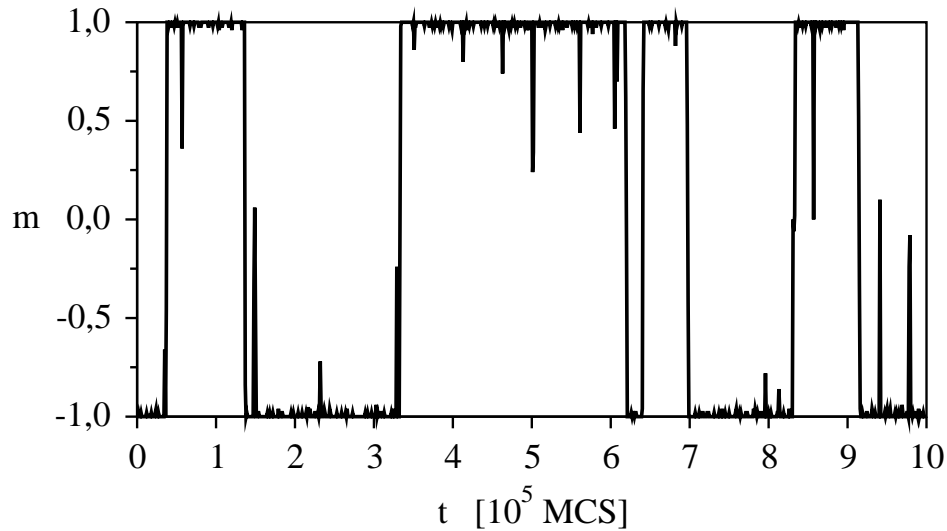}
\caption{The time evolution of the public opinion $m$ in the system of $L=100$ individuals as a function of time for the probability of anticonformity $p_1=0.003$. It can be seen that society for most of the time is in a consensus state ($m = \pm 1$), but from time to time spontaneous reorientations occur. From the social point of view this means that on the one hand society polarizes to given opinion due to the conformity, but on the other hand spontaneous (and rather rapid) changes of polarization are possible, due to the weak anticonformity.}
\end{center}
\end{figure}

\begin{figure}
\begin{center}
\vspace*{10mm}
\includegraphics[scale=1]{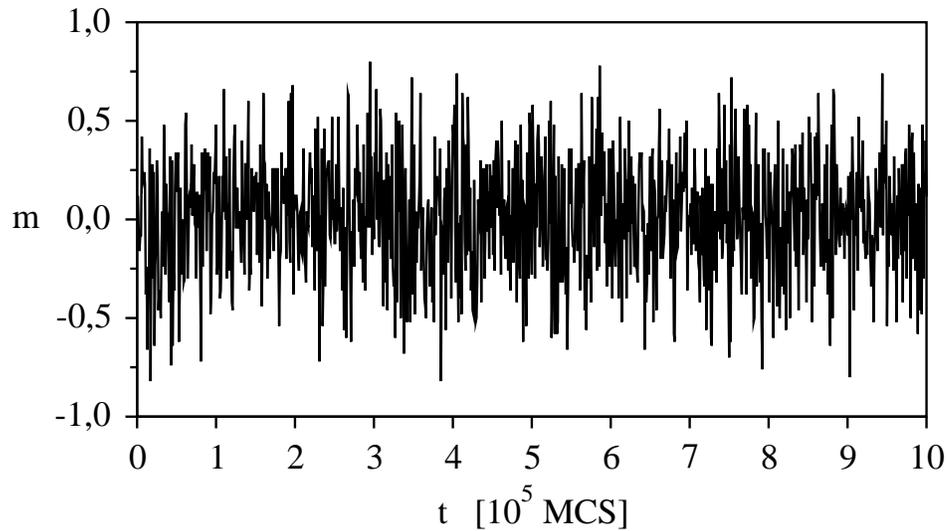}
\caption{
The time evolution of the public opinion $m$ in the system of $L=100$ individuals as a function of time for the probability of anticonformity $p_1=0.1$. It can be seen that already for this level of anticonformity consensus is not reached and the public opinion oscillates around its mean value $m=0$.}
\end{center}
\end{figure}

\begin{figure}
\begin{center}
\vspace*{10mm}
\includegraphics[scale=1]{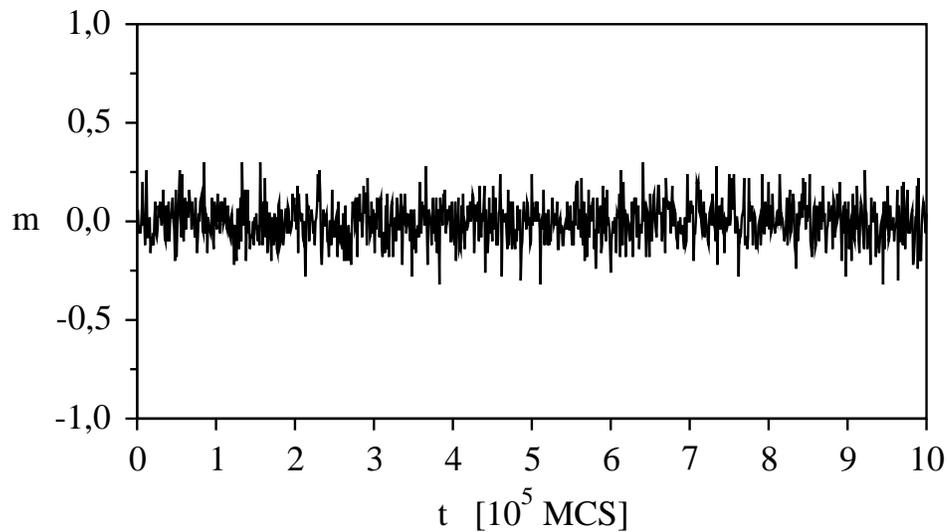}
\caption{
The time evolution of the public opinion $m$ in the system of $L=100$ individuals as a function of time for the probability of anticonformity $p_1=0.9$. It can be seen that for this level of anticonformity consensus is not reached, similarly to the Fig.2.
The difference between the case $p_1=0.1$ and $p=0.9$ is visible in the fluctuations around the mean value $m=0$ -- the amplitude of the fluctuations decreases with $p_1$, on the other hand the frequency of fluctuations increases with $p_1$.}
\end{center}
\end{figure}

To analyze more precisely the dependence between the consensus time and the level of anticonformity $p_1$ let us introduce the mean relative time of consensus $<\tau_c>$ as a mean number of $MCS$ for which $|m|=1$ divided by the total number of steps in the simulation. The dependence between the mean relative time of consensus $<\tau_c>$ and $p_1$ is presented in Fig.4. For small values of $p_1$ this dependence is exponential, i.e $<\tau_c> \sim \exp(\alpha p_1)$, with $\alpha=\alpha(L) \sim \frac{3}{2}L$. This means that although the relative time of consensus decrease with $p_1$, consensus is still possible for larger values of $p_1$. No qualitative change of behavior is seen while looking at $<\tau_c>$ as a function of anticonformity. On the other hand, if we look at Figs. 1-3 it seems that there is some qualitative difference between opinion dynamics presented in Fig.1 and Fig.2-3. In Fig. 1 the system is ferromagnetically ordered for most of the time and spontaneous transitions between two opposite ferromagnetic states are observed.

\begin{figure}
\begin{center}
\vspace*{10mm}
\includegraphics[scale=1]{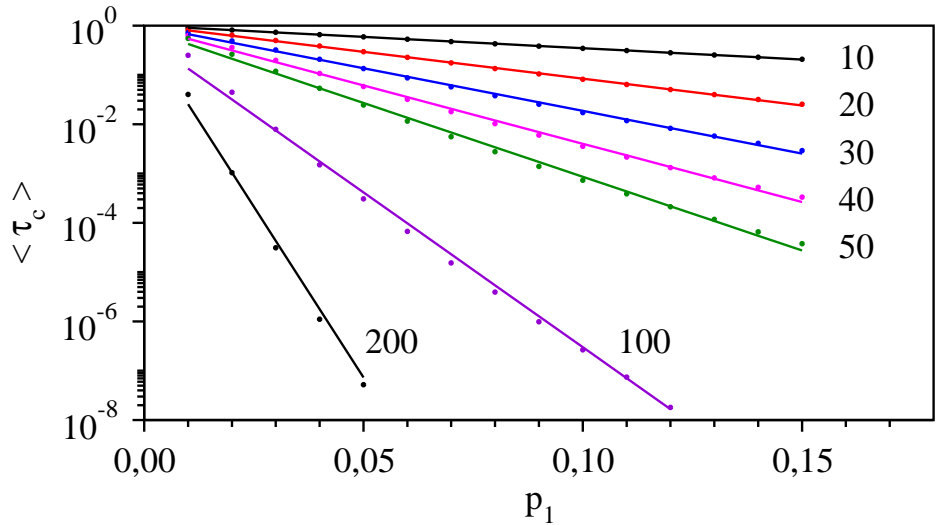}
\caption{The dependence between the mean relative time of consensus $<\tau_c>$ and the level of anticonformity $p_1$ for several lattice sizes (from $L=10$ to $L=200$). For small values of $p_1$ this dependence is exponential, i.e $<\tau_c> \sim \exp(\alpha p_1)$, with $\alpha=\alpha(L) \sim \frac{3}{2}L$.}
\end{center}
\end{figure}

Therefore, let us now check the dependence between control parameter $p_1$ and the mean reorganization time $<t_r>$, 
defined as a mean time between arriving at two consecutive opposite consensus states. More precisely we monitor the events of time, at which the system attains the given consensus ($m=\pm 1$) for the first time since it was in the last opposite state $m=\mp 1$. It occurs that there is an optimal value of $p_1$ for which the mean reorganization time $<t_r>$ is the shortest (see Fig.5). From the social point of view this means that there is a special level of anticonformity for which reorganizations (`revolutions') are the most frequent. The optimal value of $p_1$ is roughly inversely proportional to the system size $L$. Thus their product $p_1L$, describing the mean number of acts of anticonformity per one Monte Carlo step, remains constant independently on the system size.

Now we can show that indeed there is a qualitative change in the opinion dynamics for a certain value of $p_1$ and this value corresponds to the optimal value of $p_1$, i.e. value for which the mean reorganization time $<t_r>$ is the shortest. To do this let us present the cumulative distribution function $CDF$ of the public opinion $m$. In Fig. 6 it can be seen that for $p_1 \le 0.04$ the curve is $\sim$ shaped and for certain value $p_1=p^* \in (0.03,0.04)$ the shape of $CDF$ changes qualitatively to the $\backsim$ shape (the change in convexity). While for $p_1 \le 0.04$ the system for most of the time is in the consensus state, for $p_1 \ge 0.03$ the consensus state is extremely low probable. One should notice (see Fig.5) that the optimal value of $p_1$ also lies in the interval $(0.03,0.04)$ and thus corresponds to $p^*$. 

\begin{figure}
\begin{center}
\vspace*{10mm}
\includegraphics[scale=1]{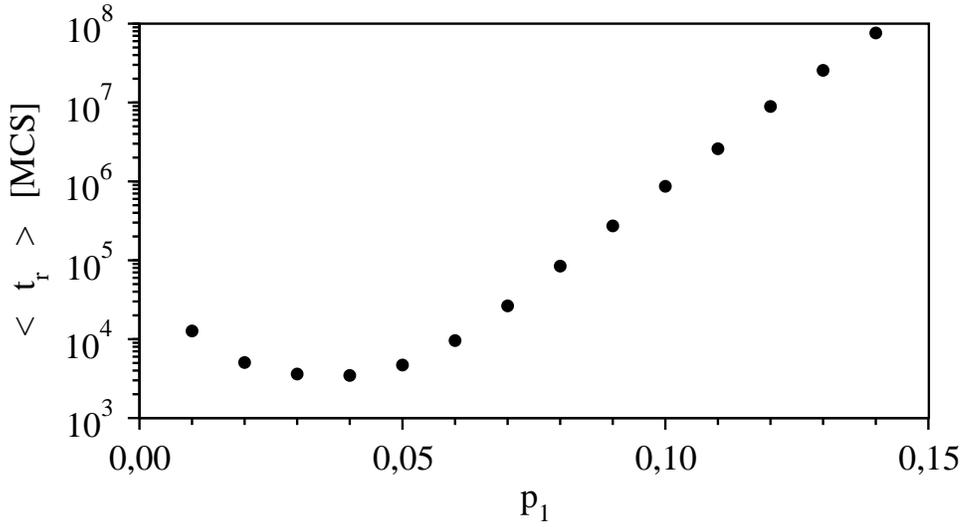}
\caption{The dependence between the mean reorganization time $<t_r>$ and the level of anticonformity $p_1$ for the lattice size $L=100$.}
\label{fig1}
\end{center}
\end{figure}

\begin{figure}
\begin{center}
\vspace*{10mm}
\includegraphics[scale=1]{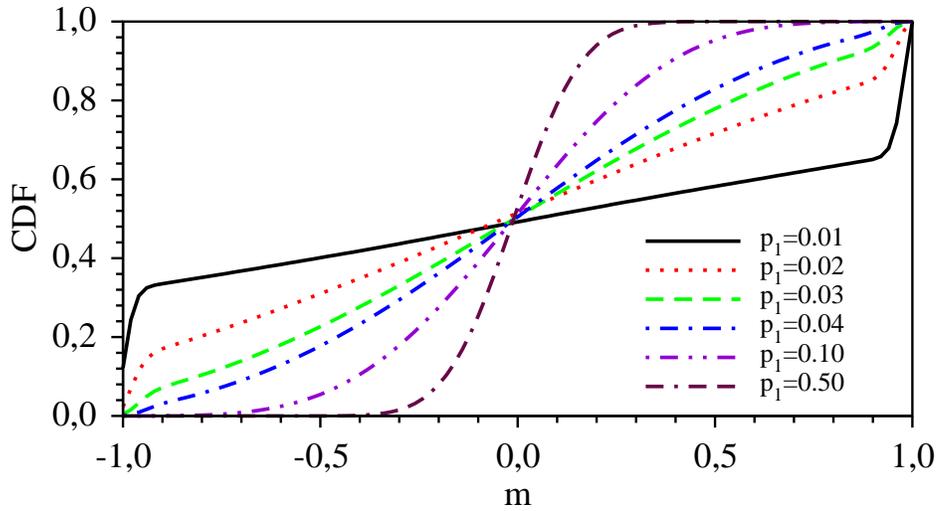}
\caption{The cumulative distribution function $CDF$ of the public opinion $m$ for several values of anticonformity level $p_1$ and the lattice size $L=100$. It can be is seen that for $p_1 \le 0.04$ the curve is $\sim$ shaped and for certain value $p=p^* \in (0.03,0.04)$ there is the qualitative change in convexity to the $\backsim$ shape.}
\end{center}
\end{figure}

\section{Summary}
We have proposed a new model of opinion dynamics with anticonformists based on the general model proposed by Kondrat \cite{Kondrat_2009}. In our model only conformity (with probability $1$) and anticonformity (with probability $p_1$) can lead to the opinion change. According to Willis, both conformers and anticonformers are similar in the sense that both acknowledge the group norm (the conformers agree with the norm, the anticonformers disagree). In our model a pair of neighboring individuals sharing the same opinion will influence its neighborhood (so called outflow dynamics -- the idea taken from the Sznajd model).  To investigate the model, we have provided Monte Carlo simulations with the random sequential updating mode. It occurs that for small values of anticonformity level consensus is still reached, but it is not the absorbing steady state as in the case of $p_1=0$. For small values of $p_1$ spontaneous reorientations occur, which can be understood from the social point of view, as complete repolarizations (e.g. spontanous transition from dictatorship to democracy).  We have shown that there is a special value of anticonformity level $p_1=p^*$ below which the system stays for most of time in the consensus state and spontaneous reorientations occur. Above this value the consensus it almost impossible and qualitative change is visible in the cumulative distribution function of the public opinion $m$. 

The main criticism connected with such simple social models concerns usually oversimplifications of the assumptions. We do not want to convince anybody that there is no free will or no external factors influencing individual choices. We have only shown that even in the conformistic societies with very low (but nonzero) level of anticonformity, spontaneous reorientations of the public opinion are possible. There is no need to introduce any external field nor strong leader to explain these social repolarizations. This seems to be quite important result in the social perspective. Sociologists usually try to explain a posteriori such a rapid and unexpected transitions (like protests, revolutions, etc.) and having known the history they are quite often able to do so. On the other hand maybe from time to time there is no direct reason for such a reorientation, maybe it occurs just spontaneously because the society is the complex dynamical system.

\end{document}